\documentclass[manuscript]{aastex}
\usepackage{lscape}
\usepackage{color,soul}
\pdfoutput=1

\received{Feb. 21, 2016}
\accepted{July 25, 2016}
\slugcomment{To appear in The Astrophysical Journal}

\shorttitle{Sub-DLA at $z=5$}
\shortauthors{Morrison et al.}

\begin{document}

\title{Element Abundances in a Gas-rich Galaxy at $z=5$: Clues to the Early Chemical Enrichment of Galaxies\footnote { Based on observations obtained at the W.M. Keck 
Observatory, which is operated as a scientific partnership among the California Institute of Technology, the University of California and the National Aeronautics and 
Space Administration. The Observatory was made possible by the generous financial support of the W.M. Keck Foundation. }}

\author{Sean Morrison}
\affil{University of South Carolina, Dept. of Physics and Astronomy, Columbia, SC 29208}

\author{Varsha P. Kulkarni}
\affil{University of South Carolina, Dept. of Physics and Astronomy, Columbia, SC 29208}
\email{kulkarni@sc.edu}

\author{Debopam Som}
\affil{University of South Carolina, Dept. of Physics and Astronomy, Columbia, SC 29208; 
Aix Marseille Universit\'e, CNRS, Laboratoire d'Astrophysique de Marseille, UMR 7326, 13388, Marseille, France}

\author{Bryan DeMarcy}
\affil{University of South Carolina, Dept. of Physics and Astronomy, Columbia, SC 29208}

\author{Samuel Quiret}
\affil{Aix Marseille Universit\'e, CNRS, Laboratoire d'Astrophysique de Marseille, UMR 7326, 13388, Marseille, France}

\author{Celine P\'eroux}
\affil{Aix Marseille Universit\'e, CNRS, Laboratoire d'Astrophysique de Marseille, UMR 7326, 13388, Marseille, France}

\begin{abstract}
Element abundances in high-redshift quasar absorbers offer excellent probes of the chemical enrichment of distant galaxies, and can constrain models for population III and early population II stars. Recent observations indicate that the sub-damped Lyman-alpha (sub-DLA) absorbers are more metal-rich than DLA absorbers at redshifts 0$<$$z$$<$3. 
It has also been suggested that the DLA metallicity drops suddenly at $z$$>$4.7. 
However, only 3 DLAs at $z$$>$4.5 and none at $z$$>$3.5 have ``dust-free'' metallicity measurements of undepleted elements. We report the first quasar sub-DLA metallicity measurement at $z$$>$3.5, from detections of undepleted elements in high-resolution data
for a sub-DLA at $z$=5.0. We obtain fairly robust abundances of C, O, Si, and Fe, using lines outside the Lyman-alpha forest. This absorber is metal-poor, with O/H]=-2.00$\pm$0.12, which is $\gtrsim$4$\sigma$ below the level expected from extrapolation of the trend for $z$$<$3.5 sub-DLAs. 
The C/O ratio is 1.8$^{+0.4}_{-0.3}$ times lower than in the Sun. 
More strikingly, Si/O is 3.2$^{+0.6}_{-0.5}$ times lower than in the Sun, while Si/Fe is nearly (1.2$^{+0.4}_{-0.3}$ times) solar. This absorber does not display a clear alpha/Fe enhancement. Dust depletion may have removed more Si from the gas phase than is common in the Milky Way interstellar medium, which may be expected if high-redshift supernovae form more silicate-rich dust. C/O and Si/O vary substantially between different velocity components, indicating spatial variations in dust depletion and/or early stellar nucleosynethesis (e.g., population III star initial mass function). The higher velocity gas may trace an outflow enriched by early stars. 
\end{abstract}

\keywords{galaxies: abundances-- quasars: absorption lines}

\section{Introduction}
Understanding the production of the chemical elements in galaxies is vital to understanding the evolution of galaxies. In fact, the cosmic odyssey of the elements is fundamental to nearly all aspects of science, including the evolution of galaxies (e.g., Pei et al. 1999, Pagel 2009), the formation of planets, and the development of life. Indeed, planets have been detected around stars older than even 11 Gyr (Silva Aguirre et al. 2015). Understanding the early phases of metal production in the universe is thus crucial for a number of reasons. 

Especially interesting are the galaxies at redshifts $z \gtrsim 5$, an epoch spanning the first 
$\sim 1.2$ Gyr 
since the Big Bang. This epoch included the formation of the population III stars and early generations of population II stars. Many recent theoretical studies have investigated the formation of metals in these early stars (e.g., Maio \& Iannuzzi 2011; Wise et al. 2012; Maio \& Tescari 2015). These studies predict that nucleosynthesis by the early stars should give rise to rapid metallicity evolution and peculiar interstellar element abundance patterns in galaxies at $z \gtrsim 5$. Supernova explosions of population III stars can also contribute a significant amount of dust production in young galaxies (e.g., Marassi et al. 2015). Measuring the element abundances of galaxies at redshifts $z \gtrsim 5$ is, however, very challenging, since observations of emission from these galaxies are difficult. 

Absorption lines in quasar spectra offer a promising tool to measure abundances in distant galaxies. The damped Lyman-alpha (DLA, $N_{\rm H I} \ge 2 \times 10^{20}$ cm$^{-2}$, Wolfe et al. 1986, 2005) and sub-damped Lyman-alpha (sub-DLA,  $10^{19} \le N_{\rm H I} < 2 \times 10^{20}$ cm$^{-2}$, P\'eroux et al. 2003) absorbers are especially important for this purpose. A primary reason for this is that DLAs and sub-DLAs allow accurate measurements of the H I column densities from the damping wings of the Lyman-alpha absorption lines. DLAs have H I column densities $N_{\rm H I} \ge 2 \times 10^{20}$ cm$^{-2}$ and sub-DLAs have H I column densities in the range $10^{19} \le N_{\rm H I} < 2 \times 10^{20}$ cm$^{-2}$ (P\'eroux et al. 2003). Given these large H I column densities, DLAs and sub-DLAs dominate the H I content of the universe, and therefore play an important role in the evolution of gas and stars in galaxies (e.g., Prochaska \& Wolfe 2009; Zafar et al. 2013; Popping et al. 2014). 

The majority of past studies of element abundances in quasar absorbers focused on DLAs, since owing to their large H I column densities, DLAs were expected 
to be associated with star formation and chemical enrichment (e.g., Nagamine et al. 2004, Pettini 2004, Wolfe et al. 2005). It was thus surprising that recent studies of sub-DLAs at $z < 3$ showed the sub-DLAs to be more 
metal-rich on average than the DLAs (e.g., P\'eroux et al. 2006; Meiring et al. 2007, 2009a,b; Prochaska et al. 2006; Som et al. 2013, 2015; Quiret et al. 2016). 
Indeed, {\it super-solar} metallicity sub-DLAs are found even at $2 < z < 3$ (e.g., Som et al. 2013). The high metallicities are not an artifact of ionization, since 
the ionization corrections for most sub-DLAs are relatively modest ($\la 0.2$ dex; e.g., Dessauges-Zavadsky et al. 2003; Meiring et al. 2007; Som et al. 2015). 
Thus, sub-DLAs appear to have undergone metal enrichment earlier and may be associated with more massive galaxies than DLAs (e.g., Kulkarni et al. 2010). 
This raises the question of whether sub-DLAs were more metal-rich than DLAs even at higher redshifts, and whether sub-DLAs with super-solar metallicities exist 
at even $z>4$. At high enough redshifts, both sub-DLAs and DLAs should be metal-poor, yet at $z<3$, they clearly differ in metallicity. Thus, understanding when 
DLAs and sub-DLAs started differentiating chemically is of great interest. Unfortunately, no reliable observations of quasar sub-DLA metallicities exist at $z > 4$. A few limits exist for $z \ge 4$ quasar sub-DLAs (e.g., three O limits by Prochaska et al. 2015); 
but no definitive measurements have been reported. [We note that metallicity limits based on undepleted elements also exist for a few $z \ge 5$ DLA/sub-DLAs in the host galaxies of gamma-ray bursts (GRBs; e.g., Kawai et al. 2006, Ruiz-Velasco et al. 2007, Hartoog et al. 2015), but most of these are based on low-resolution spectra.]

In recent years, some quasar DLAs at $z > 4$ have been observed (e.g., Rafelski et al. 2012, 2014). These studies suggest that the relatively gentle metallicity evolution of DLAs 
observed at $z <3$ holds back to $z \sim 4.5$ (although some other studies disagree--see, e.g., Jorgenson et al. 2013). Furthermore, Rafelski et al. (2014) reported a sharp drop in the 
DLA metallicity at $z > 4.7$. Such a drop, if real, could signal a sudden change in the nucleosynthetic processes, perhaps indicative of an influence from population 
III stars or early population II stars. 

One complication in determining element abundances for DLAs and sub-DLAs is that the refractory elements 
get depleted into interstellar dust grains, i.e., condense into the solid phase. Elements such as Fe, Ni, Cr, and to a somewhat smaller extent, Si, are more depleted  (e.g., Savage \& Sembach 1996; Jenkins 2009), so it is more difficult to determine metallicities from these elements. Reliable determination of the total (gas + solid phase) metallicities requires observations of volatile elements that do not show much depletion. The volatile elements most commonly used for determining metallicities of DLAs / sub-DLAs are S and Zn. At $z > 4$, however, the only available lines of Zn lie in the near-infrared, and thus are inaccessible to optical spectrographs. Elements such as O and C can be used if their column densities can be measured without problems of line saturation. Unfortunately, such dust-free measurements exist for only 3 DLAs at $z>4.6$. Likewise, only 2 sub-DLAs at
$z>4$ have O or S measurements, of which neither is at $z > 4.2$. Furthermore, both of these measurements are limits based on a single saturated line, rather than definitive values based on multiple lines. In fact, none of the sub-DLAs at $z > 3.5$ have reliable O or S measurements.

Here we report the first  reliable quasar sub-DLA metallicity measurement at $z >3.5$, based on detections of undepleted elements in high-resolution data for a sub-DLA at $z =5.0$.
 This paper is organized as follows: Section 2 describes the observations, data reduction, absorption line measurements, and column 
density measurements. Section 3 discusses the results and compares the chemical compositions and physical properties of the sub-DLA studied here with those of 
other sub-DLAs and DLAs presented in the literature. Section 4 summarizes our main conclusions. 

\section{Observations and Data Reduction}

The observations of the background quasar SDSSJ120207.78+323538.8 (hereafter Q1202+3235, $z_{em} = 5.2924 \pm 0.001259$ \citep{DR5}) consisted of two exposures of 7200 s each obtained with Keck HIRES on April 4, 2010 as part of the program UH018Hr (PI A. Wolfe) and two exposures of 1800 s each obtained with Keck ESI on March 22, 2009 (program U143E, PI A. Wolfe). The extracted, continuum-normalized, and combined HIRES spectra were downloaded from the KODIAQ database (O'Meara et al. 2015), and the raw ESI spectra were downloaded from the Keck Observatory Archive. The ESI spectra were reduced and extracted using ESIRedux, an IDL-based reduction package written by J. X. Prochaska. The extracted ESI spectra were split into smaller sections (typically $\sim 100-400$ {\AA} wide) around the lines of interest, and these sections were continuum-fitted using the IRAF ``CONTINUUM'' task\footnote{IRAF is distributed by the National Optical Astronomy Observatory, which is operated by the Association of Universities for Research in Astronomy, Inc., under cooperative agreement with the National Science Foundation.}. To fit the continuum in each section, we tried cubic spline polynomials typically of an order between two and six, and used the function that provided the best fit as judged from the rms of the residuals. The continuum-normalized sections from the two ESI exposures were combined for each wavelength range using the IRAF ``SCOMBINE'' task. In addition, the continuum around a few of the lines measured with HIRES was refitted using second order cubic spine polynomials in order to obtain a better continuum fit in these regions. 

\subsection{Absorption Line Measurements and Column Density Determinations}

\begin{figure}
\epsscale{.80}
\plottwo{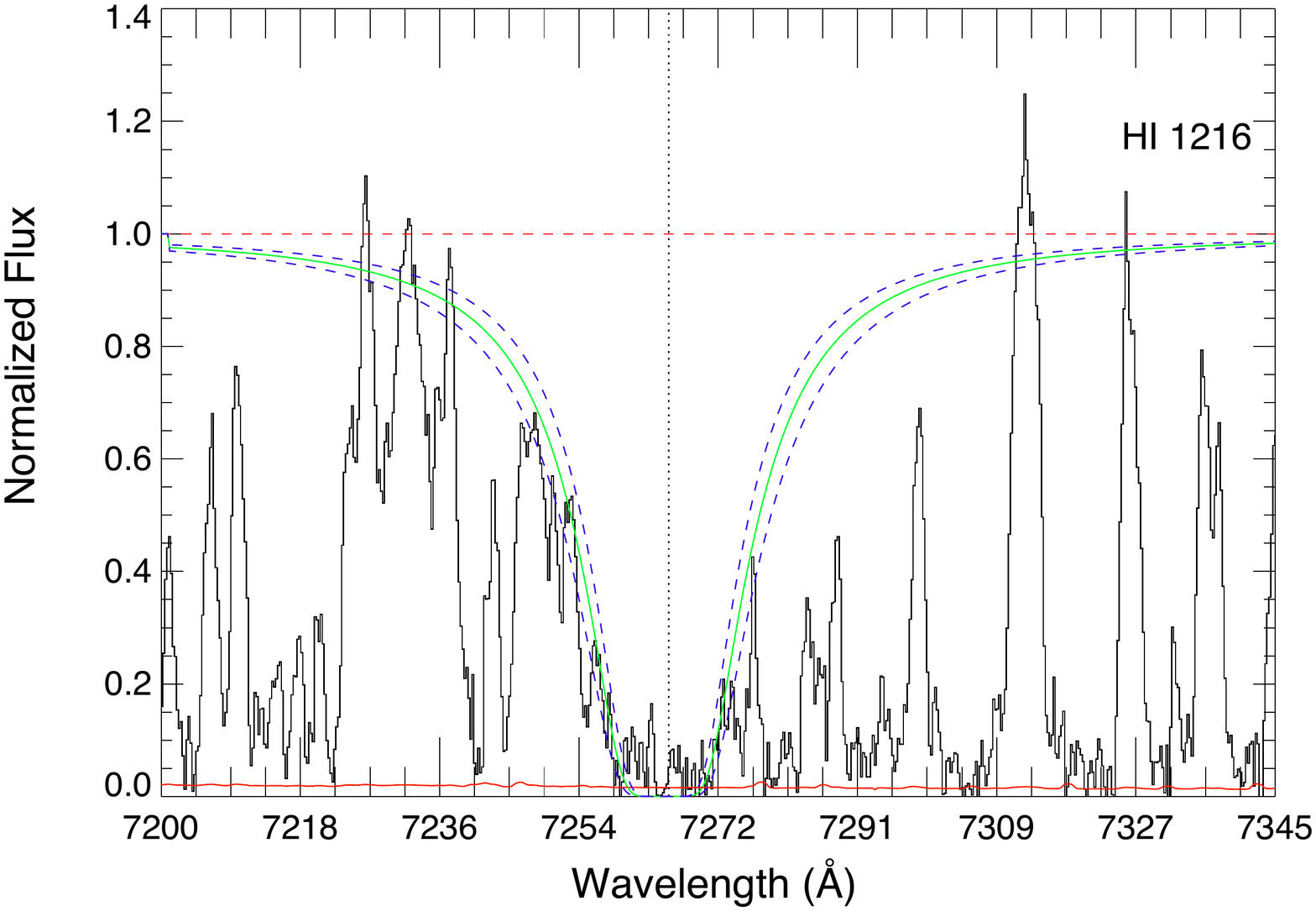}{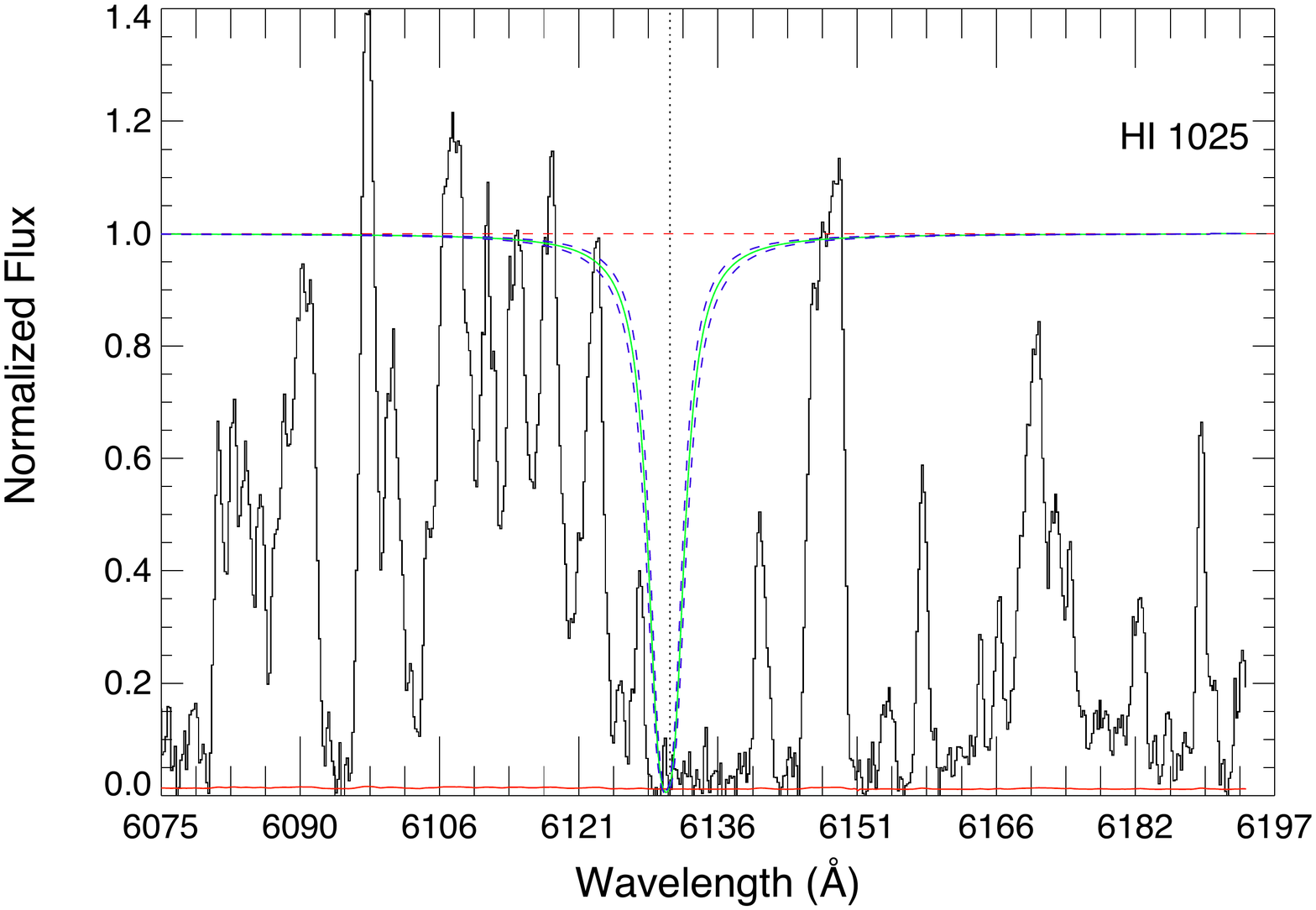}
\caption{H I Ly-$\alpha$ (left)  and Ly-$\beta$ (right) absorption features in the Keck ESI data for the $z=4.977$ absorber toward Q1202+3235. 
In each panel, the continuum-normalized flux 
is shown in black, while the $1\sigma$ error array in the 
normalized flux is shown near the bottom in red. The solid green curve represents the best-fitting Voigt profile corresponding to log $N_{\rm H I}=~$19.83 and the
 dashed purple curves denote the
profiles corresponding to estimated $\pm 1\sigma$ deviations ($\pm 0.1$ dex) from the best-fitting $N_{\rm H I}$ value. The horizontal dashed red line denotes 
the continuum level. The vertical dotted black
line denotes the center of the Ly-$\alpha$ or Ly-$\beta$ line. 
\label{LySeries}}
\end{figure}

A sub-DLA with 
$z=4.977$ was identified in the SDSS spectra of Q1202+3235 by Noterdaeme et al. (2012), who obtained log $N_{\rm HI} = 20.02$. Fig. \ref{LySeries} shows 
our Voigt profile fits to the Keck ESI data in the vicinity of the H I Lyman-alpha and Lyman-beta lines in this absorber. 
Our simultaneous fits to the Ly-$\alpha$ and Ly-$\beta$ lines give log $N_{\rm H I} =19.83 \pm 0.10$. It is clear from the left panel of Fig. \ref{LySeries}, that this $N_{\rm H I}$ value is fairly well-determined. 
Since the blends with the other Ly-$\alpha$ lines makes the task of continuum determination challenging at these high redshifts, we make use of the small stretches of continuum seen on both sides of the line in estimating the continuum level.  We note that the excess absorption seen to the left of the Ly-$\beta$ fit near 6129 {\AA} cannot be attributed to Ly-$\beta$: if this excess flux is fitted as Ly-$\beta$, then the wings of the Ly-$\alpha$ fit are correspondingly wider than observed.

Column densities were determined 
by Voigt profile fitting using the program VPFIT\footnote{http://www.ast.cam.ac.uk/ rfc/vpfit.htm} version 10.0.  When fitting the metal lines, the low ions, O I, Si II, C II, and FeII, were fitted together, with the redshifts and the Doppler $b$-parameters of corresponding components tied together. 

\begin{figure}
\includegraphics[angle=0,scale=.60]{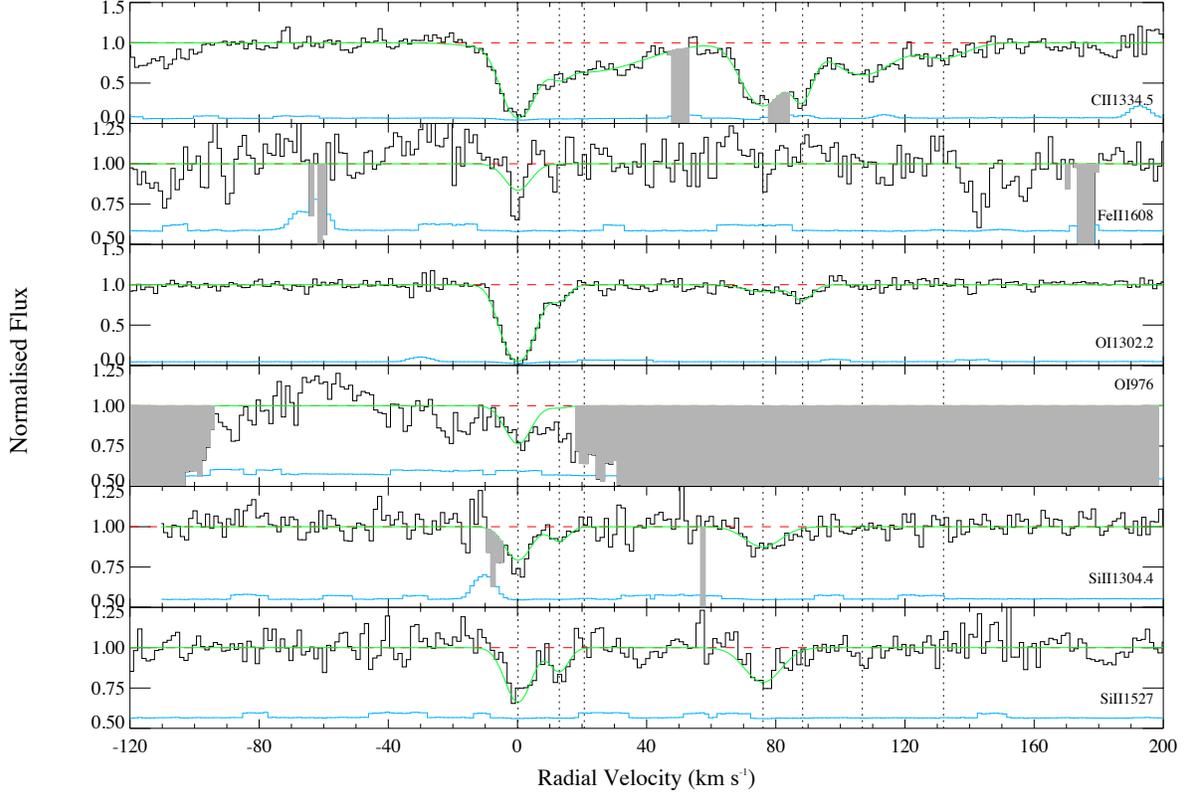}
\caption{Velocity plots from Keck HIRES data for several metal lines of interest for the $z=4.977$ system in the spectrum of Q1202+3235. In each panel, the normalized data 
are shown in black, the
solid green curve indicates the theoretical Voigt profile fit to the absorption feature, and the dashed red line shows the continuum level. 
The $1\sigma$ error values in the normalized flux
are represented by the cyan curves near the bottom of each panel. Note that in a few panels with weak lines, if the normalized flux scale shown is zoomed in, starting at 0.5, the $1\sigma$ error arrays have been offset by 0.5, so that they can be viewed in the same panels. The vertical dotted lines indicate the positions of the components that were 
used in the fits. The gray shaded regions in some of the panels represent absorption unrelated to the line presented or regions of high noise. \label{LowIons}}
\end{figure}

Fig. \ref{LowIons} shows the velocity plots for the key low-ion metal lines. 
We emphasize that all of these metal lines except O I $\lambda$ 976 lie {\it outside} the Ly-$\alpha$ forest, and thus allow accurate measurements of column densities. This fact is very important, since the primary reason for the dearth of measurements of weakly depleted elements such as O in $z \gtrsim 5$ DLAs and sub-DLAs is that their key absorption lines often lie within the Ly-$\alpha$ forest, which is very dense at these high redshifts. For C II, only one line ($\lambda$ 1334) is covered outside the Ly-$\alpha$ forest. The C II $\lambda$ 1036 line is strongly blended with the Ly-$\alpha$ forest lines, and therefore not useful. 

\begin{figure}
\includegraphics[angle=0,scale=.60]{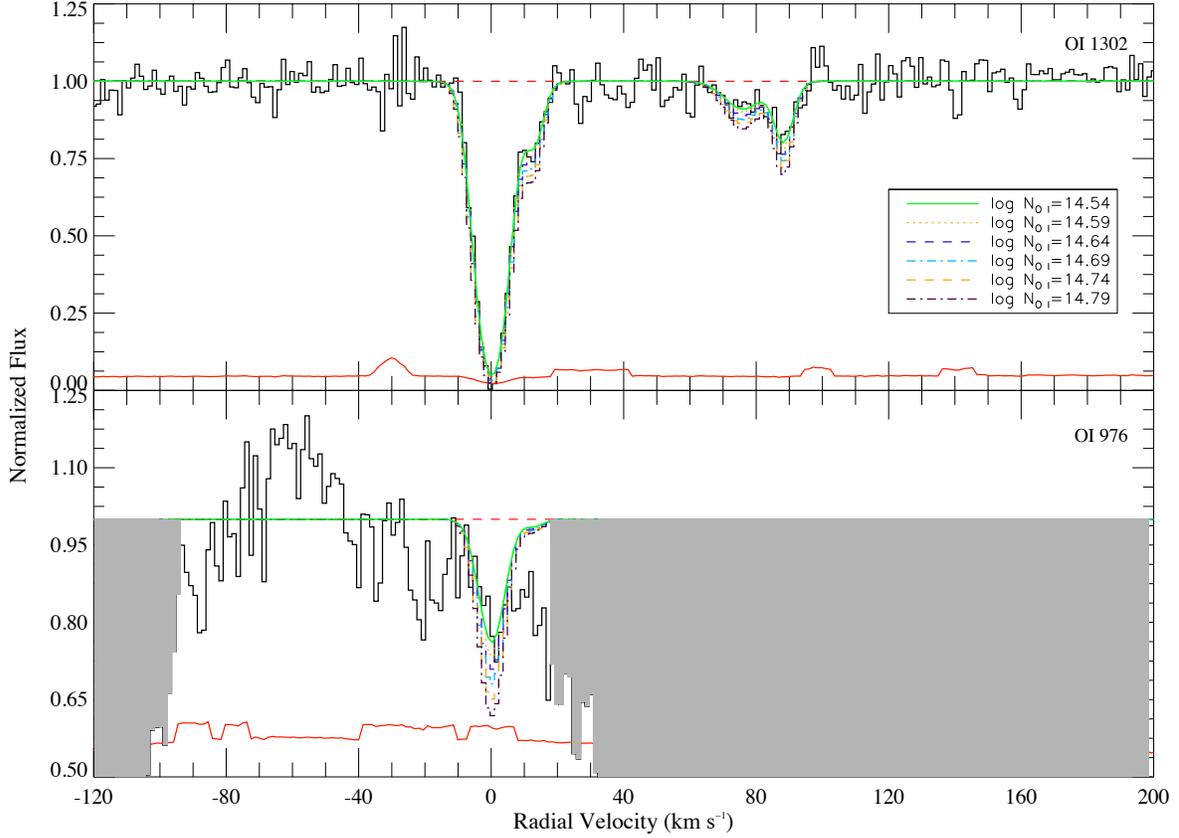}
\caption{Similar to Fig. 2, but showing the velocity plots for O I $\lambda$ 1302 and 976 only. The lines shown are Voigt profile models with incremental fits, with the green line showing the quoted total column density and the remaining fits in increments of 0.05 dex.  The $1\sigma$ error values in the normalized flux
are represented by the red curves near the bottom of each panel. Note that in the lower panel, the normalized flux scale shown is zoomed in, starting at 0.5, and the $1\sigma$ error array has been offset by 0.5, so that it can be viewed in the same panel. \label{OI_tests}}
\end{figure}

While O I $\lambda$ 1302 can often be saturated in lower redshift DLA/sub-DLAs, it does not appear to be extremely saturated in the absorber of interest here. Moreover, O~I $\lambda$ 976 is covered by the data. Even though the higher velocity components of O I $\lambda$ 976 are blended with C III $\lambda$ 977 and also the Lyman-$\alpha$ forest, the 
main component seen in O I $\lambda$ 1302 is also seen in O I $\lambda$ 976; thus we can obtain reasonable column density constraints by simultaneously fitting both the O I lines. The weaker components that are only fitted using O I $\lambda$1302 are weak enough that, even if their  fits were not perfect, the change to the total O I column density would be insignificant. Fig.  \ref{OI_tests} demonstrates what is gained by using the O I $\lambda$ 976 line by 
showing the effect of changing the O I column density in steps of 0.05 dex on the profiles of the $\lambda$ 976 and $\lambda$  1302 lines. It is clear that  even though the strongest component of O~I $\lambda$1302 changes very slowly, the corresponding component of O I $\lambda$ 976 quickly becomes too deep for the observed profile. Thus, the O I column density is unlikely to be higher than our adopted best-fit value by much more than 0.10-0.15 dex. Since the O I $\lambda$ 976 line is in the Ly-$\alpha$ forest, we also investigated the question of what the lowest O I column density consistent with O I $\lambda$ 1302 is. To do this, we examined the effect of decreasing the O I column density in steps of 0.05 dex below the best-fit value on the O I $\lambda$ 1302 profile. Comparing the corresponding profiles with the observed O I $\lambda$ 1302 profile, we determine that the O I column density cannot be lower than our adopted best-fit value by more than 0.10-0.15 dex. 
For Si II, simultaneous fitting of two lines (Si II $\lambda \lambda$ 1304, 1526, both outside the Lyman-$\alpha$ forest) allows us to obtain a robust determination of the line parameters.

For Fe II, the only available line outside the Lyman-$\alpha$ forest is Fe II $\lambda$ 1608, which is marginally detected. The stronger Fe II lines at longer wavelengths (e.g. Fe II $\lambda \, \lambda$ 2382, 2586, 2600 are not covered by the HIRES or ESI data at these high redshifts, 
while the Fe~II lines at shorter wavelengths (e.g., $\lambda \, 1144$) are strongly blended with the Lyman-$\alpha$ forest. 

\begin{figure}
\includegraphics[angle=0,scale=.60]{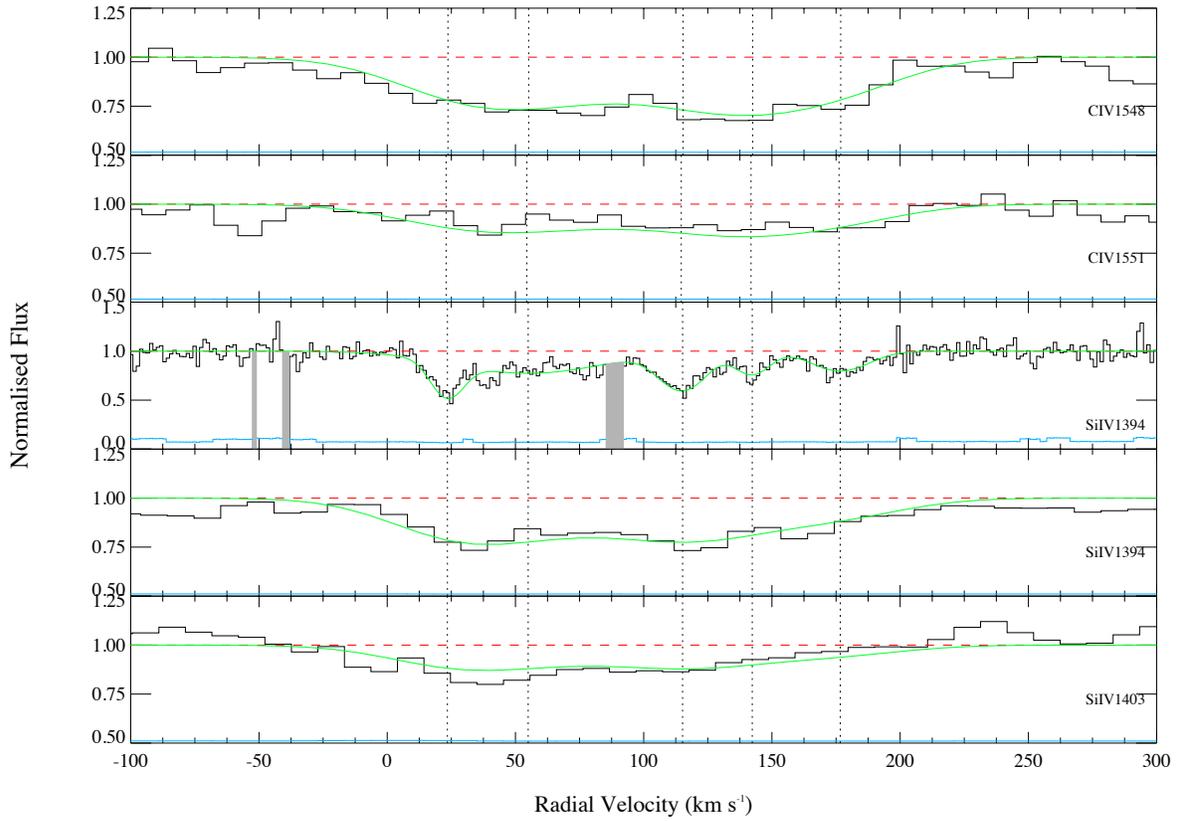}
\caption{Same as Fig. \ref{LowIons}, but for the higher ions C IV and Si IV. Data for the upper Si IV $\lambda 1394$ panel are from HIRES, while the data 
in all the other panels are from ESI. \label{HighIons}}
\end{figure}

Fig. \ref{HighIons} shows the velocity plots for the available lines of C IV and Si IV, and our Voigt profile fits for these. C IV $\lambda$ 1548, 1551 lie outside the Lyman-$\alpha$ forest, but are not covered by the HIRES data. We therefore use the ESI data for those. The Si IV $\lambda$ 1394 line is covered by the HIRES data. Si IV $\lambda$ 1403 is not covered by the HIRES data, but is covered by the ESI data. C III $\lambda$ 977 could not be fitted due to blending with Ly-$\alpha$ forest lines. Si~III $\lambda$ 1206 could not be fitted, since it is covered only in part, and is also heavily blended with Lyman-$\alpha$ forest lines. 

For fitting the high-ion metal lines, the Si IV $\lambda$ 1394 line from the HIRES data was fitted independently to obtain the velocity structure. This velocity structure was then applied to the same line in the ESI data to confirm that the fit parameters gave the same column density. After confirming the HIRES fit worked on the ESI data, the velocity structure, with the redshifts and $b$-parameters fixed, was then applied to C IV $\lambda \, \lambda$ 1548, 1551 and Si IV $\lambda \, \lambda$ 1394, 1403 from the ESI data, as well as Si IV $\lambda$ 1394 from the HIRES data in order to get the final column densities.

The green curves plotted over the data in  Fig. \ref{LySeries}-\ref{HighIons} show the Voigt profile fits to the absorption lines. 
The total 
column densities were obtained by summing over all the velocity components. These values were checked 
independently using the apparent optical depth (AOD) method (Savage \& Sembach 1991). 
For consistency with past studies (which we use to compare to our results in section 4 below), we have adopted oscillator strengths 
from Morton (2003). We note, however, that more recent oscillator strength determinations exist for a few elements (e.g., 
Cashman et al. 2016 and references therein), and 
should be uniformly applied to all future element abundance studies in DLA/sub-DLAs. 

Table 1 presents the measurements of the metal column densities for the individual velocity components for the low ions. Table 2 lists the column densities for the individual velocity components for the higher ions C IV and Si IV. 
Table 3 gives the total column densities for the available ions (summed over individual velocity components) derived from the profile fits, along with the AOD estimates. 

\section{Discussion}

\subsection{Total Element Abundances}
Table 4 lists the total element abundances for C, O, Si, and Fe calculated using the total metal column densities along with the H I column density, and using the solar abundances from Asplund et al. (2009). 
The second column of Table 4 lists the abundance values based on the dominant ions (e.g., C II for C and Si II for Si), without including ionization corrections. C appears to be modestly enriched with respect to O, with 
[C/O]$_{\rm No \, IC}$ = $ 0.21$ $\pm$ $ 0.09$. 
Fe appears to be far less abundant, with [C/Fe]$_{\rm No \, IC}$ = $0.69$ $\pm$ $0.13$, [O/Fe]$_{\rm No \, IC}$ = $0.49 \pm 0.14$, and [Si/Fe]$_{\rm No \, IC}$ = $ 0.35$ $\pm$ $ 0.12$. We next consider the effect of ionization corrections using Cloudy photoionization models. As we discuss below, 
the  moderately positive apparent value of [C/O]$_{\rm No \, IC}$ and the large apparent values of [C/Fe]$_{\rm No \, IC}$ and [Si/Fe]$_{\rm No \, IC}$ are partly due to ionization effects, since the ionization corrections are larger for C and Si than for Fe, and are essentially negligible for O.

\subsubsection{Ionization Corrections}

If the absorbing gas consists of both H I and H II regions, the ionized gas needs to be taken into account while determining element abundances. 
For DLAs, these ionization corrections are usually ignored, owing to their large H I column densities. But the ionization corrections may 
not always be small for sub-DLAs, given their somewhat lower H I column densities. Previous studies of ionization corrections in $z < 3$ sub-DLAs find them to be relatively modest (Dessauges-Zavadsky et al. 2003, Meiring et al. 2007, Battisti et al. 2012, Som et al. 2015). To assess the ionization effects in the sub-DLA toward Q1202+3235, we used photoionization 
models using the plasma simulation code Cloudy v. 13.03 (Ferland et al. 2013). First, the absorbing gas was assumed to be a slab of constant 
density, illuminated by a radiation field consisting of a mixture of an intergalactic UV background, the radiation from O/B stars, the cosmic ray 
background and the cosmic microwave background at $z_{abs}=4.977$. The UV background was adopted from Haardt \& Madau (1996) and 
Madau et al. (1999) and evaluated at $z_{abs}=4.977$. The O/B stellar radiation was based on a Kurucz model spectrum for a temperature 
$T=30,000$ K. 

\begin{figure}
\includegraphics[angle=0,scale=.63]{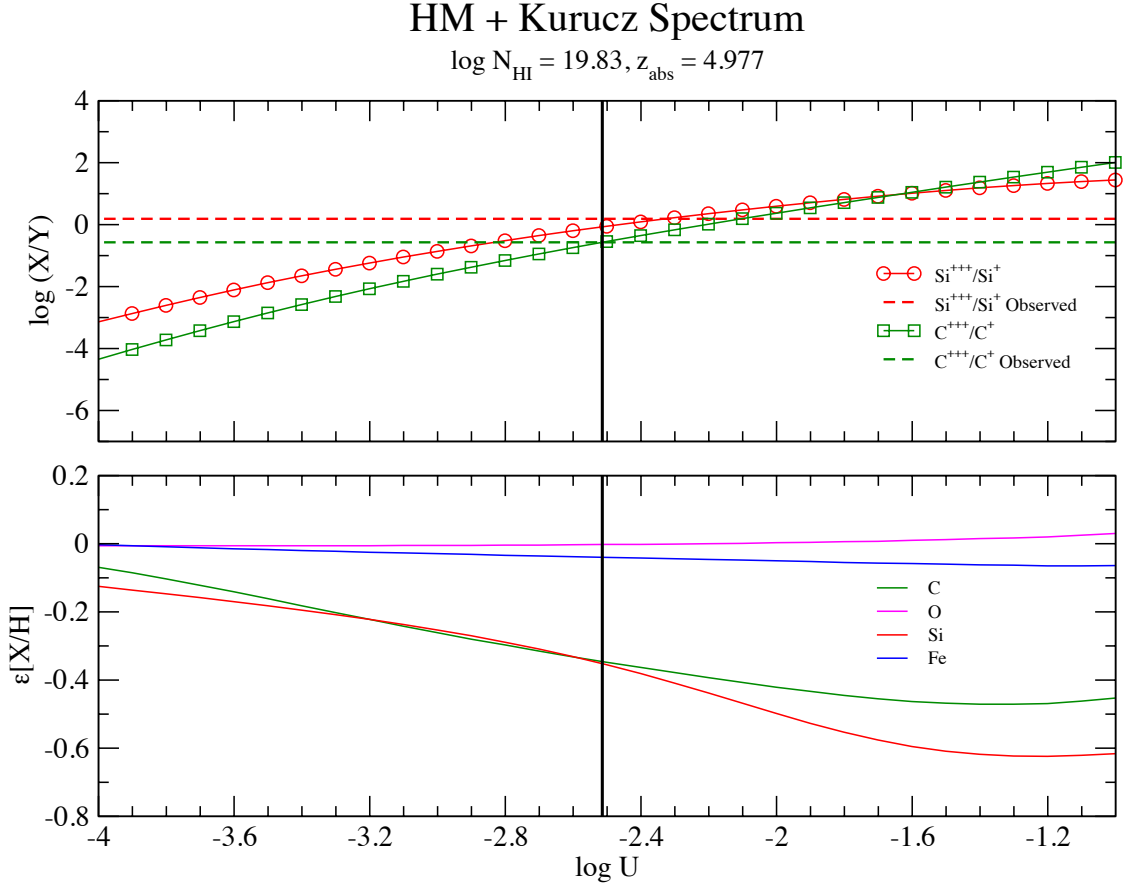}
\caption{Results of Cloudy photo-ionization calculations for the $z=4.977$ absorber toward Q1202+3235. The top panel shows the ion ratios as a function of the ionization parameter $U$. The vertical black line corresponds to the value of $U$ implied by the observed C IV/ C II ratio (indicated by the lower dashed horizontal line). The bottom panel shows the resultant estimates of the ionization corrections $\epsilon$ for the observed elements. \label{Ioniz_C}}
\end{figure}

\begin{figure}
\includegraphics[angle=0,scale=.63]{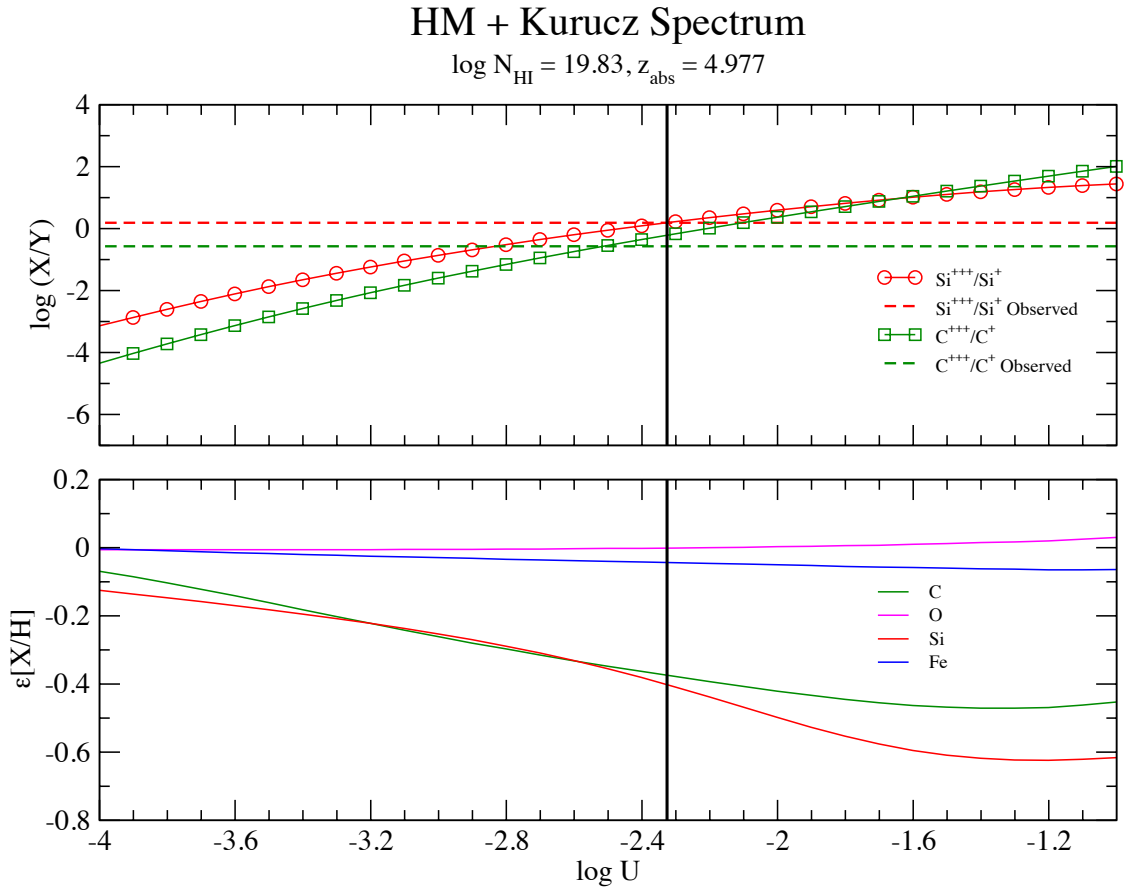}
\caption{Same as Fig. \ref{Ioniz_C}, but based on the observed Si IV/ Si II ratio (indicated by the upper dashed horizontal line). \label{Ioniz_Si}}
\end{figure}

The observed Si IV/ Si II and C IV/ C II ratios were used to constrain the ionization parameter $U = n_{\gamma}/n_{H}$, i.e. the ratio of the 
number density of ionizing photons to that of the gas. To do this, grids of models were constructed for $-6.3 <$ log $U$ $< -0.3$, followed by finer grids in the range $-4 <$ log $U$ $< -1$, 
calculating the predicted column density ratios  C~IV/ C II and Si IV/Si II (top panels of Fig. \ref{Ioniz_C} and Fig. \ref{Ioniz_Si}, respectively). The results from both these 
ratios were fairly similar. The  Si IV/ Si II constraint gave log $U$ = -2.33, while the C IV/ C II constraint gave log $U$ = -2.51. The resultant 
ionization corrections ($\epsilon$ = [X$_{\rm total}$/H$_{\rm total}$] - [X$_{\rm dominant \, ion}$/ H I]) 
are also fairly similar (bottom panels of Fig. \ref{Ioniz_C} and Fig. \ref{Ioniz_Si}). Given that the Si II and Si~IV column densities are better determined, we adopt the ionization parameter determined from 
Si IV/Si II; this implies ionization 
corrections of -0.37 dex for C, -0.001 dex for O, -0.40 dex for Si, and -0.04 dex for Fe. The corresponding ionization corrections from the C IV/ C II 
constraint agree with the above values within $< 0.05$ dex.

We next carried out Cloudy calculations without assuming a constant density, but letting the density and radiation field intensity to be varied 
simultaneously using the ``optimization'' option. This approach gave a f higher ionization parameter log $U$ = -1.44, and fairly similar 
ionization corrections for O, Si, Fe, i.e., $\epsilon$ = -0.01, -0.37, -0.10, respectively, and a somewhat different value for C, i.e., 
$\epsilon$ = -0.47. 

Columns 3, 4 and 5, 6 in Table 4 give, respectively, the abundances [X/H] and [X/O] derived using the above two sets of ionization correction estimates (``IC1'' and ``IC2'', respectively). Both determinations of the ionization corrections for O are essentially negligible ($< 0.01$ dex). This is, of course, expected, since the ionization potential of O I is very close to that of H I. We therefore regard the O abundance as the most reliable indicator of metallicity in this system. For the remaining elements, we adopt the second set of ionization corrections (``IC2'', calculated with the optimization technique). This choice was made because there is no evidence to suggest that a constant density is a realistic situation, and thus the values returned by the optimization technique are likely to be more realistic.

\subsection{Relative Abundances}

\subsubsection{Overall Abundance Pattern}

Several peculiarities are seen in the overall relative abundances. Relative to the solar ratio, C and Fe are underabundant compared to O by  0.25$ \pm 0.09$ dex and 0.58 $\pm 0.14$ dex, respectively. In other words,  the C/O ratio is 
lower than in the Sun by a factor of  1.78$^{+0.40}_{-0.32}$, while the Fe/O ratio is lower than in the Sun by a factor of 3.85$^{+1.44}_{-1.05}$. 

The most striking feature of the overall abundance pattern is that Si is significantly under-abundant relative to O, with [Si/O] of   -0.50$ \pm 0.08$. In fact Si is barely enhanced with respect to Fe,  with [Si/Fe] =  0.09$ \pm 0.12$.  Comparing the observed values [Si/C] = -0.25  $\pm  0.06$  and [Si/O] of  -0.50 $\pm 0.08$ with the corresponding values in the warm Milky Way ISM, i.e. [Si/C] = -0.22  and [Si/O] = -0.31 (Jenkins 2009), it appears that the gas-phase Si/C ratio in our $z=5$ absorber is similar to that in the Milky Way, but the gas-phase Si/O ratio is lower. 
Thus this absorber does not show a clear evidence of alpha-enrichment. This suggests that the nucleosynthetic history in this $z=5$ absorber was peculiar, and/or that Si is more preferentially depleted on to dust grains (see sections 3.4 and 4 for more discussion of these points). 

We also note that the {\it observed} column densities would, at face value, suggest a significant C enhancement relative to Fe, [C/Fe]$_{ \rm No \, IC}$ = 0.69. However, some of this excess is due to ionization effects. The ionization-corrected value is [C/Fe] =  0.33. Thus, this absorber does not show the strong C/Fe enhancement seen in some very metal-poor (VMP) DLAs observed at lower redshifts (e.g. Cooke et al. 2015) or in the C-enhanced metal-poor stars (e.g., Aoki et al. 2007). 

How does this absorber compare to the handful of DLAs studied at $z \sim 5.0$? Rafelski et al. (2014) reported measurements of Si and Fe in two DLAs at $z_{abs} > 4.9$. [Si/Fe] in these two systems are $> -0.11$ and 0.42. The [Si/Fe] in the sub-DLA reported here lies in between these values. Neither of these DLAs from Rafelski et al. (2014) have reports of O abundances, so it is not possible to compare the substantially 
negative value of  [Fe/O] =  -0.58$ \pm 0.14$ in our sub-DLA with that in other $z \sim 5.0$ absorbers. 

To help put this chemically poor $z=5$ sub-DLA in context, we also compare its relative abundances to those in the VMP DLAs at $2 < z < 4.5$ (e.g., Cooke et al. 2011). The  [Fe/O] value of -0.58 $ \pm 0.14$ in our $z=5$ absorber is somewhat low compared to the typical value of $-0.39 \pm 0.12$ for the VMP DLAs. The [C/O] value of  -0.25$ \pm 0.09$ is comparable to the typical value of $-0.28 \pm 0.12$ for VMP DLAs, while 
the [Si/O] value of -0.50 $\pm 0.08$ is much lower than the typical value of $-0.08 \pm 0.10$ for VMP DLAs (Cooke et al. 2011). Thus the $z=5$ absorber studied here appears to have had a distinct metal and/or dust production history than the lower redshift VMP DLAs.

\subsubsection{Variations in Relative Abundances} 
The bulk of the O I absorption in this absorber is limited to a much narrower velocity range, while the Si II and C II absorption is spread more uniformly. There thus appears to be a substantial difference between the relative C, O, and Si abundances between the different components. The logarithmic ratios of the observed column densities for 
C II and O I, relative to the solar C/ O ratio, are $-0.33 \pm 0.12$ dex 
and $1.03 \pm 0.12$ dex in the $z=4.977004$ and $z=4.978517$ components, respectively. The logarithmic ratios Si II/ O~I, relative to the solar Si/O ratio, are $-0.43 \pm 0.09$ dex and $1.08 \pm 0.11$ dex, respectively, in the $z=4.977004$ and $z=4.978517$ components. In other words, the observed C II/ O I and Si II/ O~I ratios in these two velocity components differ by factors of about 23 and 32, respectively. 

We cannot accurately estimate ionization corrections for these individual components seen in the low ions due to the different velocity structures of the low and high ions. However, based on the overall ionization corrections (Table 4), it is clear that the large differences between the C II/ O I and Si II/ O I ratios between these components originate at least in part in actual abundance differences. The overall under-abundance of C and Si relative to O mentioned in section 3.2.1 (obtained after summing over all velocity components) seems to be arising primarily in the main component at $z=4.977004$. By contrast, the weaker component at $z=4.978517$ (which is at a velocity of $\sim$76 km s$^{-1}$ relative to the main component) appears to have 
an enhancement of C and Si relative to O. It is possible that this higher-velocity gas traces an outflow enriched by early stars. The main component, deficient in C and Si relative to O, may be enriched by a different stellar population and may trace cooler gas with higher dust depletion. 
Such abundance differences may be further aided by ionization differences to produce the observed large differences in the C II/ O I and Si II/ O I ratios, if the outflowing higher velocity component gas is also more ionized (and hence more deficient in O I) than the main component. 

\subsection{Metallicity Evolution}

\begin{figure}
\includegraphics[angle=0,scale=.50]{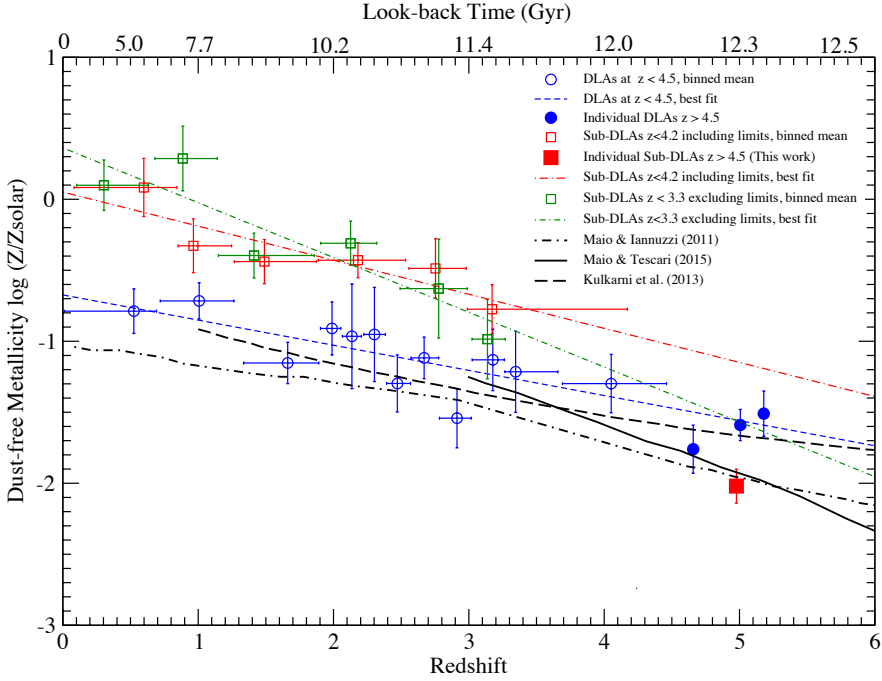}
\caption{Evolution of the dust-free metallicity traced by elements not depleted or weakly depleted on dust such as Zn, S, or O. 
Unfilled blue circles and unfilled red squares show, respectively, the binned $N_{\rm H I}$-weighted mean metallicity for DLAs at $z < 4.5$ and sub-DLAs at $z < 4.2$. Dashed blue and red lines show linear regression fits to these DLA and sub-DLA data.   Unfilled green squares show the binned $N_{\rm H I}$-weighted mean metallicity for sub-DLAs without including the limits, and the dot-dashed green line shows the corresponding linear regression fit. Filled blue circles show 
data for individual DLAs at $z > 4.5$ from Rafelski et al. (2012, 2014), while the filled red square shows the sub-DLA data at $z=5$ 
from this work. The solid, dashed, and dot-dashed black curves show the predictions from early chemical evolution models of Maio \& Tescari (2015), Kulkarni et al. (2013), and Maio \& Iannuzzi (2011). While the DLAs at $z > 4.5$ appear to be consistent with the expectation from the fit to the lower-redshift DLA data, the sub-DLA reported in this work is  below the expectations from the fit to the lower-redshift sub-DLA data by $\gtrsim$ 4.0 $\sigma$. \label{Z_evol}}
\end{figure}

Fig. \ref{Z_evol} compares the metallicity of the $z=5$ sub-DLA toward Q1202+3235 with the metallicities of other sub-DLAs and DLAs. 
The unfilled squares show the binned $N_{\rm H I}$-weighted mean metallicity vs. median redshift for the $z<4.2$ sub-DLAs , inferred from the elements Zn, S, or O that do not deplete much on interstellar dust grains, from Som et al. (2015), Prochaska et al. (2015) , Fox \& Richter (2016), Srianand et al. (2016), Quiret et al. (2016), and references therein. We have excluded absorbers along BAL sightlines and those from metallicity-selected studies (e.g., Prochaska et al. 2006, Berg et al. 2015). We note that this literature sample of sub-DLA 
metallicities is very sparse at high redshifts, consisting of only a few limits and no reliable measurements at $z > 3.3$. The majority  of the measurements are for S or Zn, since reliable measurements of O abundance based on multiple lines are available for very few systems. (Usually, O I $\lambda$ 1302 is the only line available, and is saturated.) Since the sub-DLA sample is relatively small,  we did the calculations in several different ways, both including the limits with survival analysis using the Kaplan-Meier estimator for a censored distribution, following the prescription of Kulkani \& Fall (2002), and excluding the limits. The red unfilled squares show the binned $N_{\rm H I}$-weighted mean metallicity vs. median redshift for sub-DLAs, obtained including the limits as well as detections (treating the upper limits as detections and the lower limits with survival analysis). The red dot-dashed line shows the linear regression fit to the binned sub-DLA trend thus obtained. The green unfilled squares show the trend obtained including only detections (excluding all upper and lower limits); the green dot-dashed line shows the corresponding linear regression fit. Calculations were also performed treating the upper limits with survival analysis and excluding the lower limits; the results from this last set of calculations are roughly similar to those from excluding the limits at both the low and high-redshift ends; these results are not shown in Fig. 7 to avoid cluttering the Figure, but are also mentioned below.  
The unfilled blue circles show the binned $N_{\rm H I}$-weighted mean metallicity vs. median redshift for DLAs at $z<4.5$ based on Zn, S, or O data from Rafelski et al. (2012), Guimaraes et al. (2012), Kulkarni et al. (2012, 2015), Som et al. (2015), Noterdaeme et al. (2015), Krogager et al. (2016), Srianand et al. (2016), Quiret et al. (2016), and references therein. The dashed  blue line indicates the linear regression fit to {\bf this} binned trend for the DLAs. 
The filled blue circles show the individual DLAs at $z > 4.5$ with measurements of undepleted elements (based on S) from Rafelski et al. (2012, 2014). The filled red square shows the O measurement for the sub-DLA at $z=5$ from this paper. 

The black curves show the predictions from early chemical evolution models. The solid curve shows the mean gas metallicity from the cosmological hydrodynamic computations of  Maio \& Tescari (2015). The dashed and dot-dashed black curves show,  respectively, the metallicity in star-forming regions in the hydrodynamic simulations of Maio \& Iannuzzi (2011) and the average gas-phase metallicity in the semi-analytic model of Kulkarni et al. (2013). The Maio \& Tescari (2015) models do not include population III stars, while the Maio \& Iannuzzi (2011) models do include population III stars in the range 100-500 $M_{\odot}$. The Kulkarni et al. (2013) model includes population III stars in the range 100-260 $M_{\odot}$, but is dominated by population II star formation, and thus not very sensitive to the population III model. 

A few things are striking about these comparisons:

(1) The $z=5$ sub-DLA lies below the extrapolation from the lower redshift sub-DLAs, by 7.1 $\sigma$ taking the upper limits as detections and treating the lower limits with survival analysis. The difference is 4.0 $\sigma$ if the upper limits are treated with survival analysis, and the lower limits are excluded.  Even if the 
sub-DLA trend is calculated excluding all the upper and lower limits, the  $z=5$ sub-DLA lies 3.6 $\sigma$ below 
the extrapolation from that trend.  In fact, this sub-DLA also lies 3.7 $\sigma$ below even the level expected at $z=5$ from the trend for $z < 4.5$ DLAs. Given that this sub-DLA was randomly selected based on only its high redshift, it appears that at these early epochs, 
sub-DLA abundances may have been low. 
These results, together with the $> 10$ times higher mean metallicity of  lower redshift sub-DLAs, suggest that sub-DLA metallicity may have built up rapidly between $z \sim 3.5$ and $z \sim 5$. Of course, robust measurements (not just limits) of S and O abundances are needed for many more $z > 3.5$ sub-DLAs to verify if they are also metal-poor and if so, to determine how rapid the metallicity rise was. 

(2) The low metallicity of our $z=5$ sub-DLA is consistent with both the Maio \& Tescari (2015) model,  which does not include population III stars, and  the Maio \& Iannuzzi (2011) model, which does include population III stars. This suggests that the left-over 
signatures of population III enrichment may not be significant at $z =5$. Our observed value is lower than the prediction from the model of Kulkarni et al. (2013); however, the predicted trend is sensitive to the delay time involved in the metal enrichment of the interstellar 
gas by supernovae (Kulkarni et al. 2014). More gradual enrichment may give better agreement with the metallicity observed at $z=5$. Future measurements of metallicities and relative element abundances in more sub-DLAs and DLAs at $z \ge 5$ can potentially help to discern more definitively between different chemical evolution models. 

 We note that none of these three models agree with the lower redshift sub-DLA data, even though two of them are consistent with the $z=5$ measurement.  Additionally, we note that the metallicity evolution of a given galaxy depends on the halo mass, and the mean metallicity  evolution trends predicted by the models depend on the halo mass distribution (e.g., Kulkarni et al. 2013, 2014). Thus, if the halo mass distribution of sub-DLA host galaxies  has evolved between 
 $z=5$ and $z=0$, that may explain in part why the models do not fit the sub-DLA data well across the whole redshift range.

(3) We also note that the $z > 4.5$ DLAs appear to be consistent with the extrapolation from the lower redshift DLAs. This contrasts with the sudden drop of metallicity at $z > 4.7$ claimed by Rafelski et al. (2014). This difference stems from the fact that we have focused 
only on S and O measurements, while most of the measurements of Rafelski et al. (2014) were for Si and Fe, elements that are known to be depleted 
on dust grains in the Milky Way. It is therefore important to track down whether the difference is a result of dust depletion of Si and Fe. We stress that dust depletion cannot be ignored, given the evidence for dust at high redshift (e.g., from $z \sim 5$ sub-mm galaxies, Coppin et al. 2009, Walter et al. 2012, Casey et al. 2014). Furthermore, the supernova explosions of early stars are more likely to produce silicate dust (e.g., Cherchneff \& Dwek 2009, 2010). Marassi et al. (2015) find that standard core-collapse population III supernovae can produce dust efficiently, in the range of 0.2-3 $M_{\odot}$, and that this dust is predominantly silicate. Given such efficient production of silicate dust, the gas in high-redshift galaxies could show stronger depletion of Si than in the Milky Way. Thus the sudden metallicity drop claimed at $z > 4.7$ could be due to higher depletion levels. 

Finally we note that a roughly comparable metallicity lower limit ($>-1.85$ dex) has been reported for a $z=5.91$ sub-DLA associated with a GRB host (Hartoog et al. 2015). The higher metallicity of the latter system compared to our $z=5$ quasar sub-DLA may be due to a higher-than-average star formation rate in the GRB host galaxy.

\subsection{Dust Depletion}
A comparison of the abundances of refractory and volatile elements offers a tool to estimate the extent of dust depletion (e.g., Savage \& Sembach 1996, Jenkins 2009). Jenkins (2009) 
proposed a method based on multiple elements for assessing the extent of dust depletion and the determination of true (undepleted) 
abundances. We now attempt to determine the level of dust depletion in the $z=5$ sub-DLA toward Q1202+3235 applying the prescription of 
Jenkins (2009) to element abundances available from our data, as described in a study of absorbers from the ESO UVES advanced data products quasar sample 
(Quiret et al. 2016). 

\begin{figure}
\includegraphics[angle=0,scale=.80]{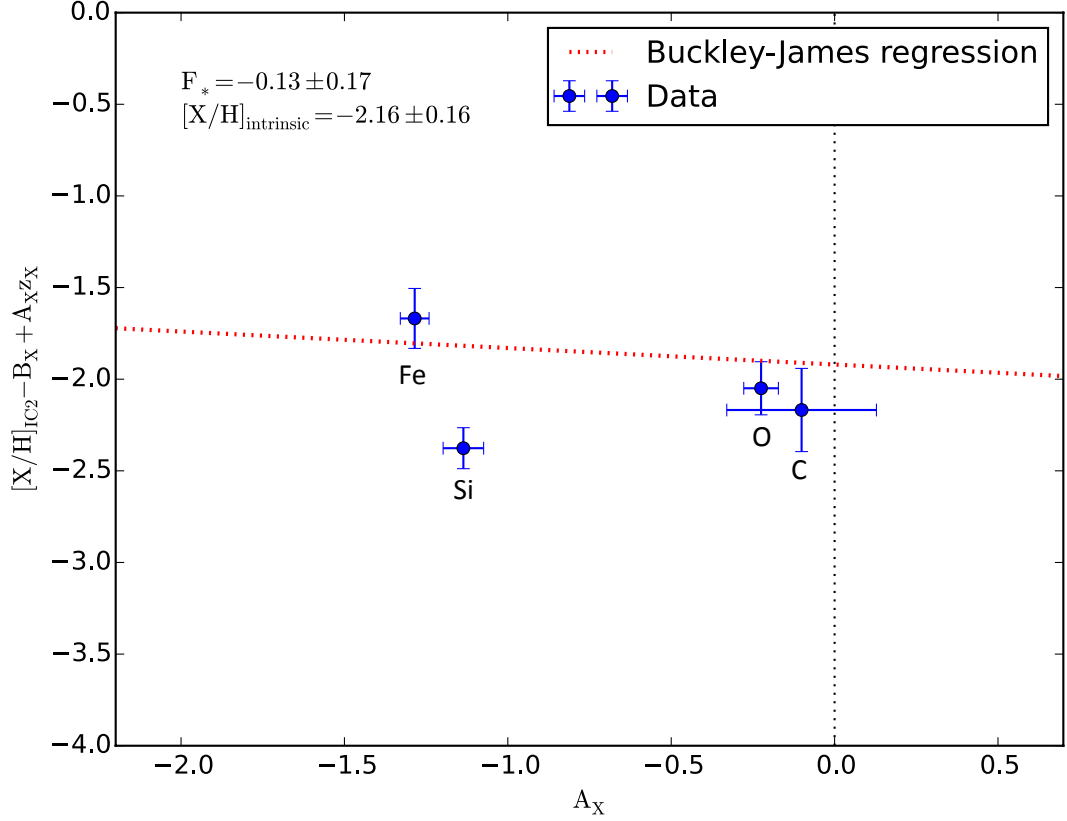}
\caption{ Determination of $F_{*}$ and intrinsic metallicity using equation A5 of Quiret et al. (2016) based on the prescription of Jenkins (2009). 
The filled circles denote the data points. The tilted dotted line shows the Buckley-James fit to the data points. The vertical dotted line corresponds to 
A$_{\rm X} = 0.0$. \label{Dep}}
\end{figure}

Fig. \ref{Dep} shows a plot of the quantity [X/H]$_{\rm IC2} -$ B$_{\rm X}$ + A$_{\rm X}$z$_{X}$ vs. A$_{\rm X}$ for the four elements available for 
this absorber. Here [X/H]$_{\rm IC2}$ are the abundance measurements corrected for ionization using the optimization technique (from column 4 of Table 4). 
The parameters A$_{\rm X}$, B$_{\rm X}$, and z$_{X}$ are adopted from Table 4 of Jenkins (2009). The dotted line shows the Buckley-James regression fit to 
the data. The $y$-intercept of this line gives the intrinsic (undepleted) metallicity, while its slope is the depletion factor $F_{*}$. Using a bootstrap 
technique with 1000 iterations,  we obtain [X/H]$_{\rm intrinsic}$ = -2.16 $\pm$ 0.16 and $F_{*}$ = -0.13 $\pm$ 0.17. The intrinsic metallicity thus inferred is consistent with 
the metallicity we obtain from O (-2.00 $\pm$ 0.12). The $F_{*}$ value is a little 
higher than the average value of $-0.34 \pm 0.19$ for the larger sub-DLA sample from Quiret et al. (2016). This suggests that the sub-DLA at 
$z=5$ toward Q1202+3235 may be somewhat more dusty than the typical lower redshift sub-DLAs, perhaps further supporting the possibility that the depletion levels may be higher at $z=5$. For reference, representative $F_{*}$ values are 
-0.1 for the warm ionized interstellar medium (ISM) of the Milky Way, 0.1 for the warm neutral medium, and 0.4 for the cold neutral medium (Draine 2011). 
The $z=5$ sub-DLA toward Q1202+3235 may be more similar to the warm ionized ISM, which is not entirely surprising. (We point out that Quiret et al. 2016 
find the typical $F_{*}$ value for DLAs to be even more negative $-0.70 \pm 0.06$, again suggesting warm ionized gas and lower dust content 
compared to the Milky Way ISM.) It would be interesting to obtain $F_{*}$ values for the high-$z$ DLAs as well. More detections of undepleted elements in high-$z$ DLAs are needed for this purpose, since none of the high-$z$ DLAs observed so far have enough such detections to permit determination of $F_{*}$. 
 
\subsection{Gas Kinematics}

Another important indicator of the absorber properties can be obtained from the velocity structure of the absorption lines. A correlation between the 
gas velocity dispersion and metallicity has been observed for both DLAs and sub-DLAs 
(e.g., Ledoux et al. 2006; Meiring et al. 2007; Som et al. 2015). If the velocity dispersion is determined primarily by the mass of the host galaxy, 
the velocity-metallicity correlation may indicate a mass-metallicity relation. Alternately, 
the velocity dispersion may be determined primarily by turbulent motions, outflows, or inflows of gas. 

\begin{figure}
\includegraphics[angle=90,scale=.30]{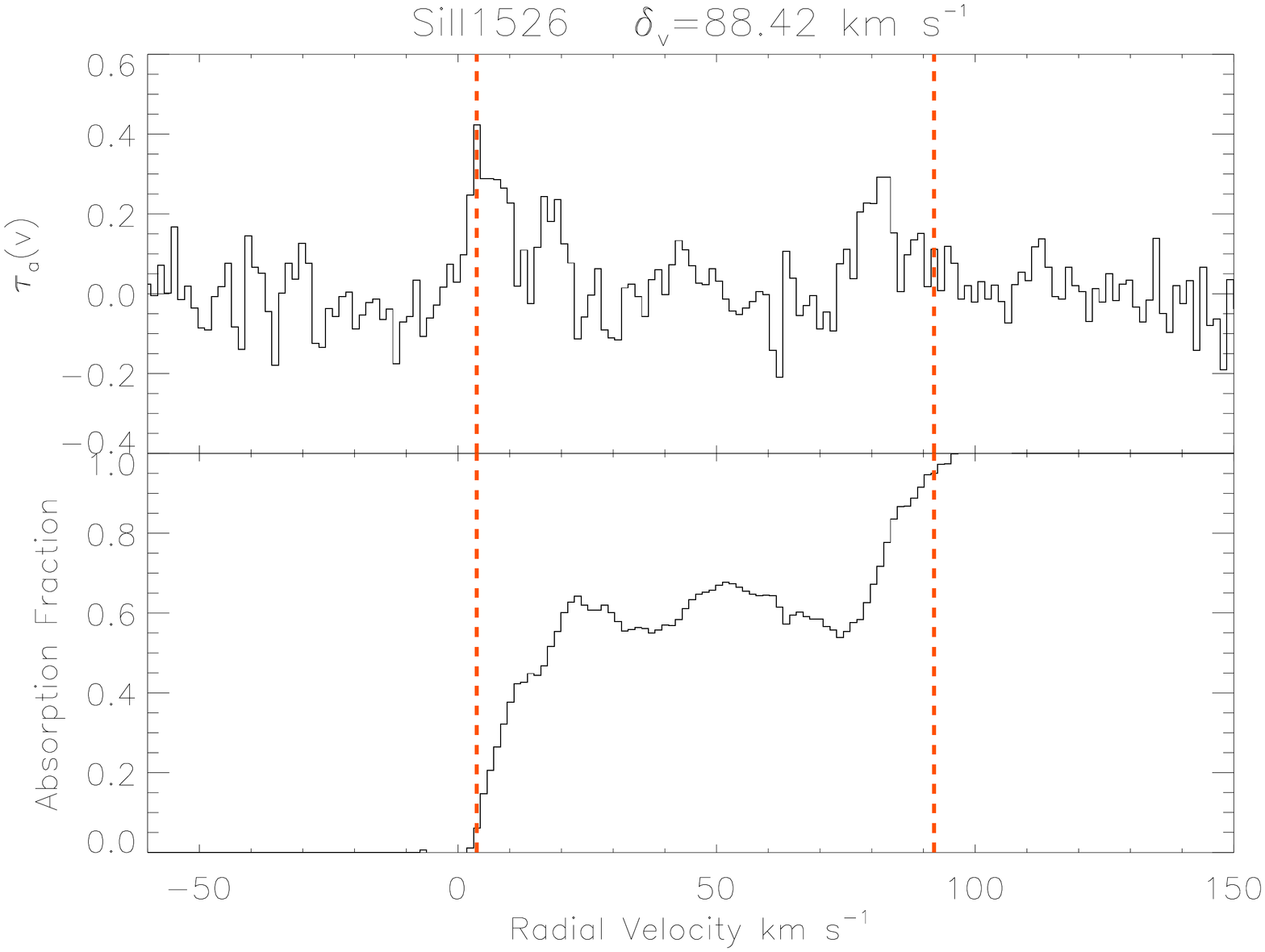}
\includegraphics[angle=90,scale=.30]{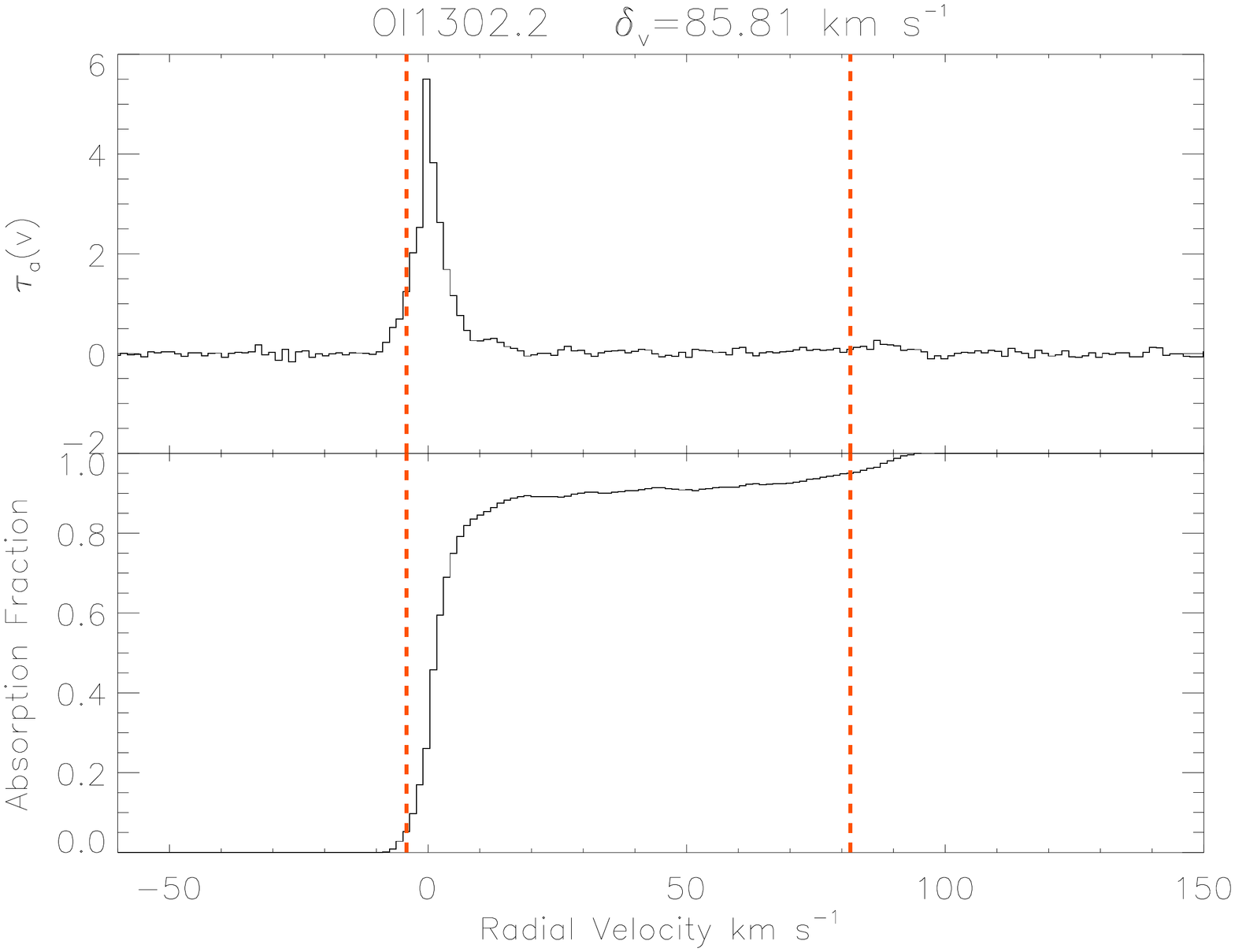}
\includegraphics[angle=90,scale=.30]{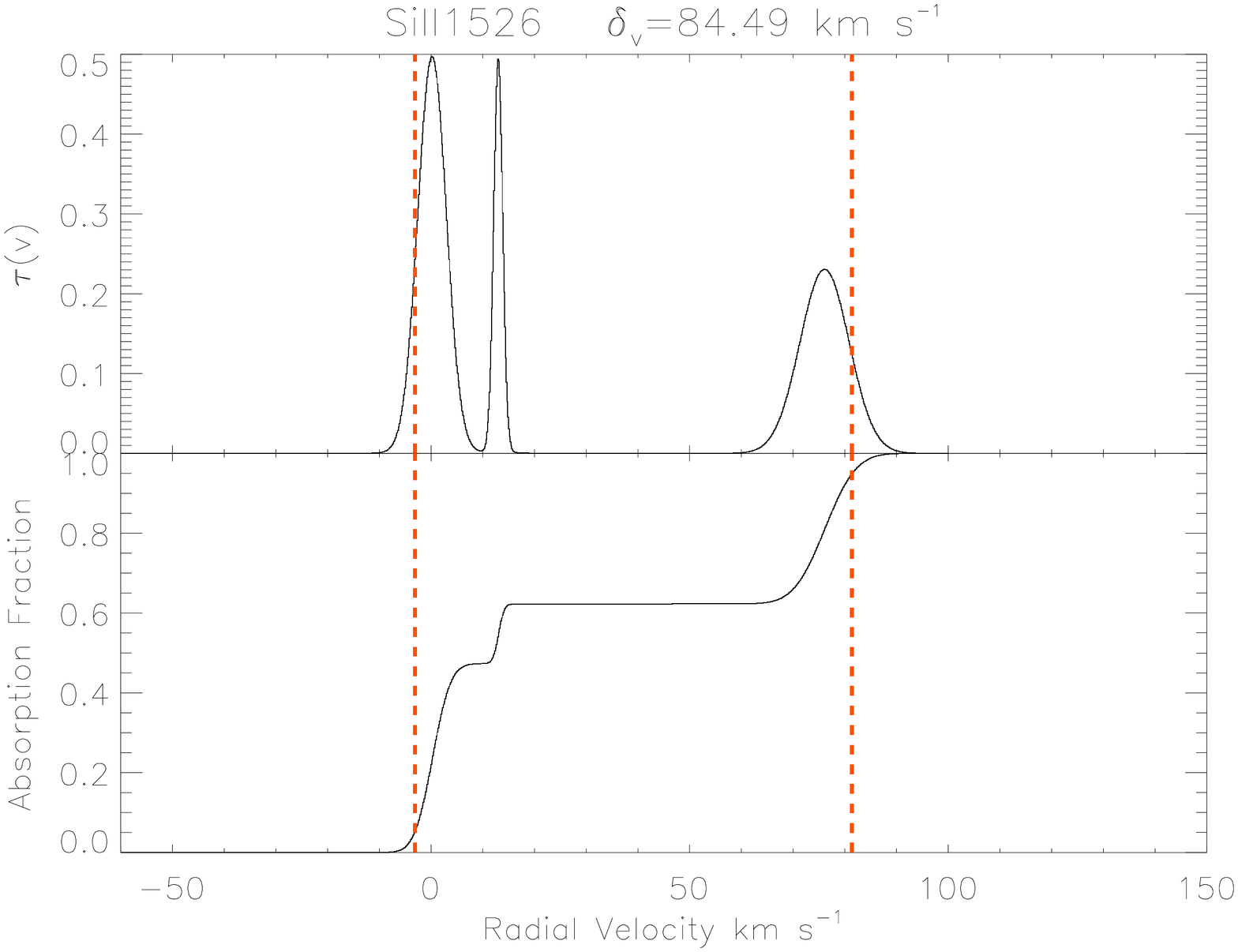}
\includegraphics[angle=90,scale=.30]{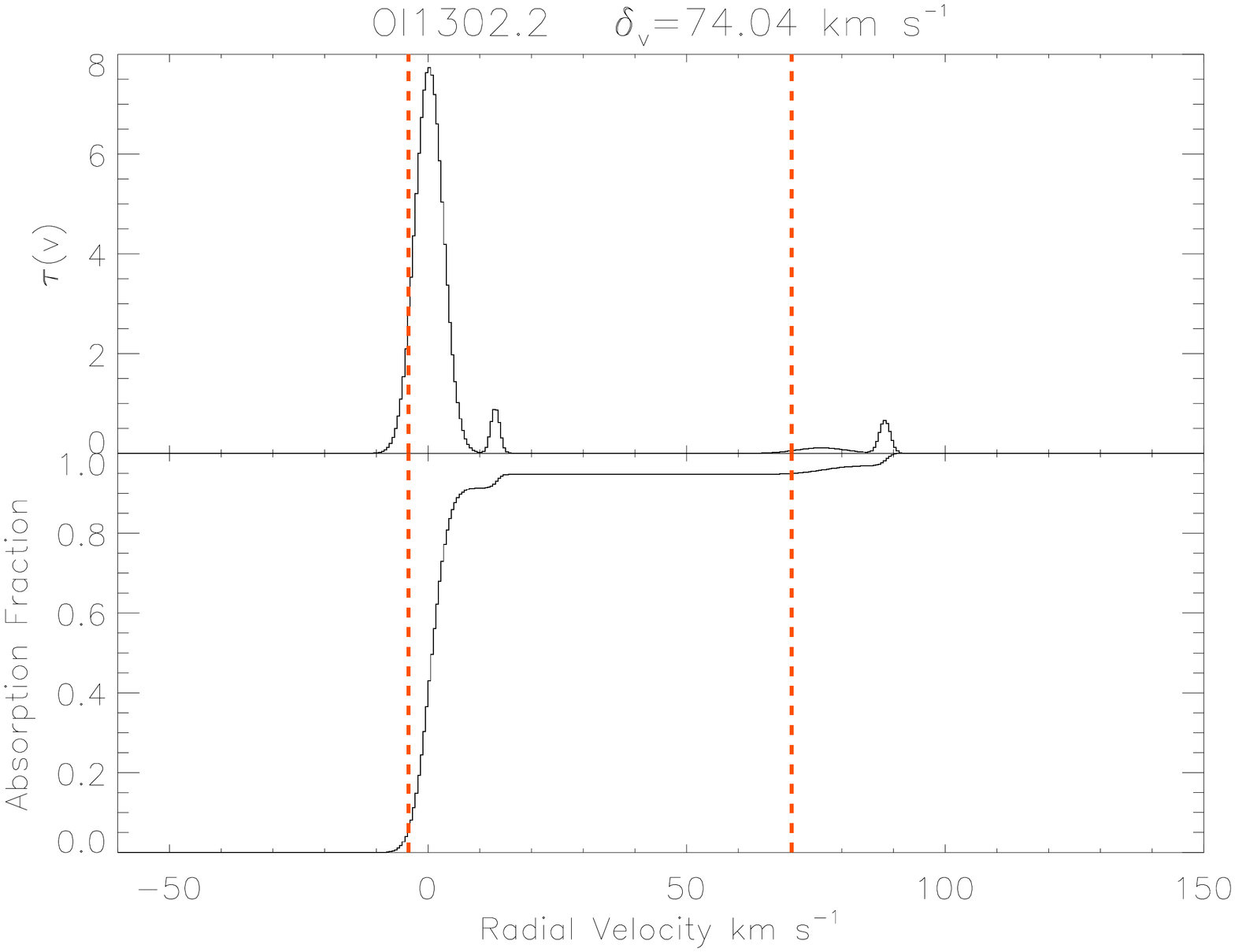}

\caption{Estimates of the velocity dispersions for the O I 1302 and Si II 1526 lines in the $z=5$ sub-DLA toward Q1202+3235, based on the observed data (upper two panels) and the theoretical Voigt profile fits (lower two panels). The top and bottom sub-panels 
for each panel show the optical depth profile and the absorption fraction as functions of the velocity of the 
absorbing gas relative to $z=4.977$. 
The red dashed vertical lines mark the 5\% and 95\% levels of absorption. \label{Dv}}
\end{figure}

The velocity-metallicity relation appears to be different for 
DLAs and sub-DLAs (Som et al. 2015). It is therefore interesting to compare the gas kinematics of the $z=5$ sub-DLA studied here with that for 
other sub-DLAs and for DLAs. The velocity width $\Delta v_{90}$ is determined as the velocity range within which 90\% 
of the integrated optical depth is contained. The top two panels of Fig. \ref{Dv} show the measurements of $\Delta v_{90}$ for Si II $\lambda$ 1526 and O I $\lambda$ 1302, based on our observations. Despite the difference in the strengths of the lines, 
the velocity dispersions inferred from their observations are quite similar, i.e., 88.42 km s$^{-1}$ for Si II $\lambda$ 1526 and 85.81 km s$^{-1}$ for O I $\lambda$ 1302. 
Quiret et al. (2016) recently advocated using the theoretical Voigt profile fit without convolution with the instrumental profile instead of the actual data for calculating $\Delta v_{90}$. The idea is 
to reduce the sensitivity of the measurement to instrumental resolution and to the presence of noise or saturation in the data. This latter 
approach, illustrated in the bottom two panels of Fig. \ref{Dv}, gives $\Delta v_{90}$ of 84.49 and 74.04 km s$^{-1}$, respectively, for the Si II $\lambda$ 1526 and O I $\lambda$ 1302 lines. Since the Si~II $\lambda$ 1526
line is less saturated, we adopt its $\Delta v_{90}$ as the velocity dispersion for this system. 

Given this velocity dispersion, this absorber lies far below the metallicity-velocity relation observed for lower redshift sub-DLAs. For example, Som et al. (2015) 
obtain [X/H] = ($0.89 \pm 0.08$) log $\Delta$ $v_{90} - (1.90 \pm 0.16$) for lower redshift sub-DLAs. As per this relation, one would expect the $z=5$ sub-DLA toward 
Q1202+3235 to have a metallicity of $-0.19 \pm 0.22$ dex, far above the observed value of $-2.02 \pm 0.12$ dex. For DLAs, Som et al. (2015) estimate 
[X/H] = ($1.02 \pm 0.03$) log $\Delta$ $v_{90} - (3.10 \pm 0.06$). Even as per this relation, an absorber with a velocity dispersion of 84.49 km s$^{-1}$ would be expected to have 
a metallicity of -1.13 $\pm 0.08$ dex. The absorber toward Q1202+3235 thus appears to lie closer to the velocity-metallicity relation for DLAs than that for sub-DLAs, and in fact, lies substantially below even the DLA relation. We do note again, however, that the bulk of the O I absorption in this absorber is limited to a much narrower velocity range. The Si II and C II absorption is spread more uniformly. This difference may arise from a difference in abundances and/or ionization in the higher velocity, possibly outflowing gas.

We also note that the velocity dispersion of the Si II gas in this absorber is much larger than the velocity dispersion measured for the very metal-poor DLAs by Cooke et al. 2015 (even after accounting for their use of the innermost 68 $\%$ rather than the innermost 90 $\%$ of the optical depth to define the velocity dispersion). Cooke et al. (2015) have suggested that the very metal-poor DLAs arise in galaxies with low halo masses. 
It appears that the $z=5$ sub-DLA studied here may be larger in halo mass. On the other hand, if the higher velocity component represents outflowing gas, and the velocity dispersion of the main absorber is in fact smaller, the implied halo mass could be smaller. 

\section{Summary and Outlook for Future Work}

We have studied the chemical composition and kinematics of a sub-DLA at $z=5$, providing the first reliable measurement of element abundances based on detections of undepleted elements in a quasar sub-DLA at $z > 3.5$. This absorber is substantially metal-poor ([O/H]= $-2.00 \pm 0.12$), $\gtrsim$ 4 $\sigma$ lower than the expectations from the $z < 3.5$ sub-DLAs. It has unusual relative abundance patterns, e.g., an overall under-abundance, relative to the solar ratio, of Si and C compared to O by factors of $\sim$3.2 and $\sim$1.8, respectively. The relative abundance pattern is also distinct from the typical pattern seen in the VMP DLAs. Furthermore, there appears to be a substantial variation in the C/O and Si/O ratios between different velocity components. Indeed, the component offset by $\sim 76$ km s$^{-1}$ from the dominant component appears to have significant enhancement of C and Si relative to O. It is possible that this higher-velocity gas traces an outflow enriched by early stars. The main component, deficient in C and Si relative to O, may be enriched by a different stellar population, or may be reflecting a higher level of dust depletion. 
With a gas velocity dispersion of $\sim$84 km s$^{-1}$, this absorber lies closer to the velocity-metallicity trend observed for DLAs than that for sub-DLAs, and in fact lies substantially below even the DLA relation. The possibility of higher Si depletion in this $z=5$ absorber also suggests that the sudden metallicity drop reported in the DLA metallicity at $z > 4.7$ could perhaps be caused in part by higher dust depletion levels. In any case, the low [Si/O] ratio in this absorber underscores the need to obtain measurements of undepleted or weakly depleted elements such as O in other DLAs and sub-DLAs at these high redshifts. 

Our results, if confirmed by studies of other sub-DLAs at similar redshifts, would suggest that sub-DLA metallicity may have built up rapidly between $z \sim 3.5$ and $z \sim 5$. The metal-poor nature of this absorber and its location in the velocity-metallicity plane also suggest that the DLA and sub-DLA absorber populations may have been more similar at high redshifts. Indeed, an absorber that is a sub-DLA at low redshift could have been a DLA at high redshift. For example, if a subset of DLAs experienced more rapid star formation (and hence neutral gas consumption) compared to the remaining DLA population, they could later manifest themselves as the less gas-rich, but more metal-enriched sub-DLAs. A systematic study of undepleted elements in more high-$z$ DLAs and sub-DLAs is essential to understand whether the two populations were indeed similar in the past, and when they started differentiating from each other. It would be fascinating to examine, for example, whether this differentiation was linked to the emergence of the dichotomy between early-type and late-type galaxies. 

The evolution of interstellar element abundances is governed by the history of star formation and the distribution of these elements in interstellar space. The cosmic star formation history has been measured out to $z \sim 10$ (e.g., Bouwens et al. 2015 and references therein). It will therefore be very interesting to examine how the metal evolution implied by this cosmic star formation history compares with the observations of high-$z$ DLAs and sub-DLAs. 

The relative abundances of elements in high-$z$ DLAs and sub-DLAs will also offer excellent constraints on metal and dust production by population III and early population II stars. For example, it will be interesting to examine whether the low [Si/O] seen in the $z=5$ absorber toward Q1202+3235 was common at that epoch. Population III nucleosynthesis models predict substantial differences between relative abundances such as [Si/O] at high redshifts for different population III initial mass functions (IMFs) and different halo masses. [Si/O] is more constraining than [C/Fe] because the Si yield varies more with the IMF, due to the efficiency of Si production in massive stars as a result of O burning. For example, Kulkarni et al. (2013) predict a substantial under-abundance of Si relative to O at $z=6$ for a population III IMF extending over 35-100 $M_{\odot}$ stellar mass range. For that IMF, a [Si/O] ratio comparable to that seen in the sub-DLA studied here would be expected for a $\sim 10^{8.5}$ $M_{\odot}$ halo at $z=6$. On the other hand, Kulkarni et al. (2013) predict a substantial overabundance of Si relative to O for a population III IMF extending over the 100-260 $M_{\odot}$ range. The differences seen in the Si/O ratios between the velocity components of the $z=5$ sub-DLA toward Q1202+3235 thus could be indicative of spatial variations in the population III IMF. Any insights obtained into the population III IMF from more relative abundance studies in high-$z$ DLAs and sub-DLAs would also help to quantify the role played by population III stars in the reionization of the intergalactic medium. Future observations of DLAs and sub-DLAs at $z > 5$ will thus provide many crucial insights into the first 1 Gyr of galaxy evolution. 

\acknowledgments
We thank an anonymous referee for helpful comments. SM, VPK, and DS gratefully acknowledge support from the National Science Foundation grant AST/1108830 and NASA grant 
NNX14AG74G (PI Kulkarni). Additional support from NASA Space Telescope Science Institute grant HST-GO-12536.01-A and NASA Herschel Science Center grant 1427151 (PI Kulkarni) 
is gratefully acknowledged. DS also acknowledges support from the A*MIDEX project (ANR- 11-IDEX-0001-02) funded by the 
``Investissements d'Avenir" French Government program, 
managed by the French National Research Agency (ANR). SQ acknowledges CNRS and CNES support for the funding of his PhD work. CP thanks the European Southern Observatory science visitor program for support. 
The authors wish to recognize and acknowledge the very significant cultural role and reverence that the summit of Mauna Kea has always had within the indigenous Hawaiian community. 
We are most fortunate to have the opportunity to conduct observations from this mountain. 

{\it Facilities:} \facility{VLT (UVES)}, \facility{Keck (ESI)}.

\clearpage

\begin{landscape}												
\begin{table}												
\begin{center}												
\caption{Results of Voigt profile fitting for lower ions in the $z=4.977$ Absorber toward Q1202+3235}
\vskip 5pt												
\begin{tabular}{ccccccccc}												
\tableline\tableline												
$ z$					&	$b^{1}_{eff}$	&	log $N_{\rm C II}$	&	log $N_{\rm O I}$	&	log $N_{\rm Si II}$	&log $N_{\rm Fe II}$	\\	
\tableline											
$4.977004\pm0.000002$	&	$3.9\pm0.2$	&	$13.91\pm0.08$	&	$14.50\pm0.08$	&	$12.89\pm0.04$	&$12.86\pm0.12$	\\
$4.977259\pm0.000009$	&	$1.2\pm1.9$	&	$12.63\pm0.28$	&	$13.07\pm0.21$	&	$12.39\pm0.13$	&				\\
$4.977413\pm0.000039$	&	$22.6\pm1.9$	&	$13.61\pm0.04$	&	\ldots			&	\ldots			&				\\
$4.978517\pm0.000008$	&	$6.7\pm0.6$	&	$13.66\pm0.07$	&	$12.89\pm0.10$	&	$12.79\pm0.05$	&				\\
$4.978761\pm0.000008$	&	$1.5\pm0.2$	&	$14.00\pm0.12$	&	$13.02\pm0.07$	&	\ldots			&				\\
$4.979130\pm0.000022$	&	$11.3\pm 2.0$	&	$13.38\pm0.06$	&	\ldots			&	\ldots			&				\\
$4.979631\pm0.000040$	&	$8.8\pm2.9$	&	$12.90\pm0.12$	&	\ldots			&	\ldots			&				\\	
\tableline												
\tableline												
\end{tabular}												
												
\tablenotetext{1}{$b_{eff}$ denotes the effective Doppler $b$ parameter in km s$^{-1}$. Column densities are in cm$^{-2}$.}												
\end{center}												
\end{table}												
\end{landscape}												
														
\begin{table}												
\begin{center}												
\caption{Results of Voigt profile fitting for higher ions in the $z=4.977$ Absorber toward Q1202+3235}												
\vskip 5pt												
\begin{tabular}{cccccc}												
\tableline\tableline												
$ z$					&	$b^{1}_{eff}$	&	log $N_{\rm C IV}$	&	log $N_{\rm Si IV}$	\\
\tableline							
$4.97747\pm0.00001$	&	$7.6\pm1.3$	&	$12.77\pm0.19$	&	$12.62\pm0.07$	\\
$4.97810\pm0.00007$	&	$39.2\pm4.6$	&	$13.51\pm0.04$	&	$12.98\pm0.04$	\\
$4.97930\pm0.00002$	&	$11.7\pm1.4$	&	$13.03\pm0.12$	&	$12.76\pm0.05$	\\
$4.97984\pm0.00002$	&	$8.4\pm1.8$	&	$13.22\pm0.10$	&	$12.37\pm0.10$	\\
$4.98052\pm0.00003$	&	$14.8\pm2.0$	&	$13.17\pm0.06$	&	$12.51\pm0.06$	\\			
	
\tableline												
\end{tabular}																						
												
\tablenotetext{1}{$b_{eff}$ denotes the effective Doppler $b$ parameter in km s$^{-1}$. Column densities are in cm$^{-2}$.}												
\end{center}												
\end{table}																							
		
\begin{table}						
\begin{center}						
\caption{Total Column Densities for the $z=4.977$ Absorber toward Q1202+3235}					
\vskip 10pt						
\begin{tabular}{ccc}						
\tableline\tableline						
Ion	&	log $N^{fit}$		&	log $N^{AOD}$		\\
	&	(cm$^{-2}$)		&	(cm$^{-2}$)		\\
\tableline										
H I	&	$19.83\pm0.10$	&	\ldots			\\	
C II	&	$14.48\pm0.05$	&	$14.32\pm0.01$	\\	
C IV	&	$13.91\pm0.04$	&	$13.85\pm0.02$	\\	
O I	&	$14.54\pm0.07$	&	$14.61\pm0.22$	\\	
Si II	&	$13.21\pm0.03$	&	$13.12\pm0.05$	\\	
Si IV	&	$13.40\pm0.03$	&	$13.37\pm0.01$	\\	
Fe II	&	$12.86\pm0.12$	&	$12.84\pm0.10$	\\	
\tableline						
\end{tabular}						
\end{center}						
\end{table}						
				
\begin{table}						
\begin{center}						
\caption{Total Element Abundances Relative to Solar}						
\vskip 10pt						
\begin{tabular}{cccccc}						
\tableline\tableline						
Element	&	[X/H]$_{No \, IC}^{1}$&	[X/H]$_{IC1}^{2}$	&	[X/H]$_{IC2}^{2}$	&	[X/O]$_{IC1}^{2}$	&	[X/O]$_{IC2}^{2}$	\\		
		&					&					&	(Adopted)			&					&	(Adopted)			\\		
\tableline		

C 		&	 -1.78$\pm 0.11$	&	  -2.15$ \pm 0.11$	&	 -2.25$\pm0.11$		&	 -0.16$ \pm 0.09$	& -0.25$ \pm 0.09$	\\			
O 		&	 -1.99$ \pm 0.12$	&	 -1.99$ \pm 0.12$	&	-2.00$\pm0.12$ 	&	...				&	...				\\	
Si 		&	-2.13$\pm0.10$		& -2.53$ \pm 0.10$	&	-2.50$\pm 0.10$ 	&	-0.54$\pm 0.08$	&	 -0.50$ \pm 0.08$	\\		
Fe 		&	-2.48$\pm0.15$		& -2.52$ \pm0.15$ 	&	-2.58$\pm0.15$		& -0.53$\pm 0.14$	&        -0.58$ \pm 0.14$	\\

\tableline						
\end{tabular}						
\tablenotetext{1}{Abundance estimates based on the dominant metal ionization states and H I, with no ionization correction.}	
\tablenotetext{2}{Estimates including ionization corrections estimated in two different ways. See the text for definitions of IC1 and IC2. Uncertainties in ionization-corrected abundances include measurement uncertainties in the relevant column densities, but not in the ionization correction estimates.}	
					
\end{center}						
\end{table}						
\clearpage						
\end{document}